\documentclass[runningheads]{llncs}
\usepackage{graphicx}
\usepackage{amsmath}
\usepackage{booktabs}
\usepackage{multirow}
\usepackage{xcolor}
\usepackage{threeparttable}
\usepackage{subcaption}
\usepackage{array,ragged2e}
\usepackage{tabularx}
\usepackage{microtype}
\usepackage{tikz}
\usetikzlibrary{calc,arrows.meta}

\begin{document}

\title{SoK: Consensus for Fair Message Ordering}

\author{Zhuolun Li \and Evangelos Pournaras}
\authorrunning{Z. Li and E. Pournaras}
\institute{School of Computer Science, University of Leeds, Leeds, UK\\
\email{sczl@leeds.ac.uk}, \email{e.pournaras@leeds.ac.uk}}

\maketitle

\begin{abstract}
Distributed ledgers rely on consensus protocols to commit messages in an agreed order. In practice, this order is often chosen for profit, which raises fairness concerns. For example, in decentralized finance, nodes exploit transaction order to extract Maximal Extractable Value (MEV). This paper systematizes the mechanisms at the consensus layer that constrain message ordering in Byzantine fault tolerant systems. We decompose adversarial behavior into three capabilities that affect ordering and analyze each ordering family against them. We organize existing work into FIFO, random, and blind ordering, and compare the metadata and assumptions each requires. We also introduce a new formalization, the $f$-Robust Median Order, which exposes a gap in existing fairness definitions based on timestamps. Finally, we describe a reusable pattern for adding fair ordering to existing BFT consensus protocols. For the protocols we analyze, fair ordering costs 1.5 to 18 times the communication of vanilla consensus, and ordering computation varies by up to five orders of magnitude across families.
\end{abstract}

\section{Introduction}
Consensus protocols in blockchains establish agreement on the order in which messages (or transactions, used interchangeably) are processed. Smart contract applications often reach different states under different orders. Since nodes have wide freedom to censor and reorder messages, they can manipulate the application state. For example, in Bitcoin~\cite{bitcoin_whitepaper} and Ethereum~\cite{ethereum_whitepaper}, proposers favor transactions with higher fees and delay the rest, which raises fairness concerns~\cite{age-aware_fairness}.

In decentralized finance (DeFi), nodes have economic incentives to reorder transactions and extract Maximal Extractable Value (MEV), often at the expense of users~\cite{MEV_research}. For example, a node that sees a pending transaction which will raise token price can front-run it with its own purchase. MEV extraction reached around 200 million USD per year in 2022~\cite{MEV_research} and has since grown.

Existing surveys focus on the application layer. Heimbach and Wattenhofer~\cite{MEV_SoK} survey how MEV arises in DeFi rather than in consensus. Yang et al.~\cite{MEV_countermeasures_SoK} classify MEV countermeasures across layers, but do not formalize ordering rules or compare their metadata assumptions. Baum et al.~\cite{baum2022sok} systematize front-running mitigations such as commit-reveal schemes and threshold encryption. Materwala et al.~\cite{materwala2024maximal} taxonomize MEV detection and mitigation at the application layer.

Our work instead takes the perspective of the consensus layer. To our knowledge, prior surveys do not offer the following contributions.

\begin{enumerate}
    \item \textbf{Adversary capability framework.} We decompose Byzantine behavior into three capabilities that affect ordering: observation falsification, suppression, and scheduling; and analyze every protocol family against them (Table~\ref{tab:adv_capability}).
    \item \textbf{Taxonomy and practitioner guidance.} We classify proposals by ordering objective, metadata, adversary tolerance, and complexity, with comparison tables (Tables~\ref{tab:FIFO_comparison} and~\ref{tab:random_comparison}), and provide a practitioner decision matrix mapping scenarios to approaches (Table~\ref{tab:guidance}).
    \item \textbf{Formalization of median timestamping robustness.} We introduce the $f$-Robust Median Order (Definition~\ref{def: robust median order}), an analytical separation condition. It exposes a gap in existing fairness definitions based on timestamps, in which Byzantine nodes can swap the median timestamps of two messages.
    \item \textbf{Integration pattern with overhead analysis.} We describe a reusable design pattern for adding fair ordering to BFT consensus and state its safety and liveness propositions (Propositions~\ref{prop:safety} and~\ref{prop:liveness}). We also derive analytical overhead estimates. The protocols we analyze incur $1.5\times$ to $18\times$ the communication cost of vanilla consensus (Table~\ref{tab:overhead}).
\end{enumerate}

\begin{figure}
    \centering
    \includegraphics[width=0.52\textwidth]{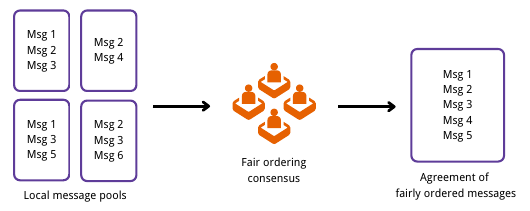}
    \caption{An example consensus protocol achieving FIFO order agreement among four nodes, committing messages in the order 1--5. Not all messages commit in one round. Message 6 stays in the local pool for a later round.}
    \label{fig:what_is_fair_consensus}
\end{figure}

The core challenge in fair ordering is that nodes do not share the same view. As Figure~\ref{fig:what_is_fair_consensus} shows, each node receives messages from users independently, so their local views of the message pool differ. Even after nodes exchange their pools, network delay and asynchrony~\cite{raynal2022communication} cause them to receive messages at different times and in different orders. Agreeing on a fair order is therefore difficult. A centralized receiver~\cite{correia2004tolerate} would avoid the problem but gives up decentralization. Most ledgers also require Byzantine fault tolerance (BFT), which further limits how robust any fair ordering rule can be.

\section{Background}

\subsection{Message Ordering and Order Sensitivity}

In a distributed ledger, nodes receive, relay, and agree on an ordered sequence of messages for execution. Each consensus round commits an ordered batch, and the order within the batch determines the state transitions. Amiri et al.~\cite{data-dependent_ordering} define \emph{order sensitive} messages as those accessing a shared resource, where the resulting state changes if the execution order changes. The widely used approach of ordering by fee addresses time sensitivity but not order sensitivity. Paying a high gas fee usually buys priority, but a later message can always pay more and jump ahead. Major distributed ledgers therefore provide no direct mechanism for protecting order sensitive messages.

\subsection{Message Ordering in Existing Distributed Ledgers}

Byzantine consensus guarantees agreement on a total order but does not constrain how that order is chosen. Li et al.~\cite{verifiable_ordering} survey ordering policies across distributed ledgers and find wide variation. Some protocols prioritize by transaction fee, others by local reception time or message origin. These policies are implementation defaults, and nodes can change them for profit. DAG based ledgers~\cite{SOK_DAG_SMR} share this weakness~\cite{mahe2025order}. Avalanche~\cite{avalance_consensus}, for example, adopts FIFO by default, but rational nodes can replace it with custom rules, and because peers cannot detect or challenge the deviation, the manipulated order remains valid under the protocol.

\subsection{Maximal Extractable Value (MEV)}

This ability to manipulate message order gives rise to Maximal Extractable Value (MEV)~\cite{MEV_SoK,MEV_analysis,MEV_research}, the profit a node extracts by controlling the sequence of transactions. MEV is especially relevant in decentralized finance (DeFi), where order influences token prices and creates arbitrage opportunities. Nodes front-run or back-run victim transactions with messages of their own~\cite{MEV_analysis}, or copy profitable pending transactions and submit imitations with higher priority~\cite{MEV_imitation_attack}. Some argue that MEV can improve market efficiency~\cite{MEV_positivity}, but it is also an attack on fairness~\cite{MEV_SoK}. Defenses at the application level exist~\cite{a2mm,order_split_prevent_sandwith}, but they do not address the root cause that the ordering mechanism can be manipulated.

\subsection{Proposer-Builder Separation and Layer 2 Sequencers}
MEV extraction has prompted major architectural shifts, most notably Proposer-Builder Separation (PBS)~\cite{proposer_builder_separation,Flashbots} and Layer 2 sequencers~\cite{sok_decentralized_sequencers}. In PBS, which currently dominates in Ethereum, block construction (by the builder) and commitment (by the proposer) are separate roles. This creates an explicit market for ordering. Builders compete to extract as much value as possible, and proposers accept the most profitable block. 

Layer 2 rollups similarly rely on a single sequencer, or a small federated committee, to order transactions. This concentrates ordering power, and fair ordering becomes crucial as rollups move to decentralized sequencers. Although our analysis uses a standard BFT model ($n=3f+1$), the vulnerabilities and mitigations we systematize apply to both PBS and decentralized L2 sequencers. We revisit fair ordering's incompatibility with PBS in Section~\ref{sec:pbs}.

\subsection{Permissioned and Permissionless Fair Ordering}
Permissioned settings assume a fixed or controlled set of participants, whereas permissionless settings allow nodes to join or leave freely. For fair ordering there is no widely accepted notion of permissionless operation. The closest is player replaceability~\cite{voting_ordering_permissionless}, in which consensus participants can be replaced without affecting security or fairness guarantees.

Many fair ordering mechanisms can be deployed permissionlessly through committee rotation or reinitialization~\cite{hybrid_consensus}. This changes the assumptions under which the guarantees hold, however. A rotating committee also depends on Sybil resistance, stake distribution, committee sampling, and repeated cryptographic setup. Repeated threshold key generation is expensive, and slow rotation gives an adaptive adversary more time to corrupt the active committee. We therefore treat permissionless deployment as an added assumption instead of a mechanical conversion from the permissioned case.

\section{System Model and Adversary Capabilities}
This section defines the scope, system model, and adversary capabilities that underpin the ordering analysis in the subsequent sections.

\subsection{Scope and Method}\label{sec:scope}

We include protocols that explicitly constrain transaction ordering at or before consensus. These are protocols whose committed output must satisfy an ordering rule, or whose step before consensus produces metadata that participants validate before voting. This covers FIFO mechanisms based on local arrival observations and random permutations applied to an agreed input set. We also discuss blind ordering, which hides transaction contents, but we classify it as an orthogonal privacy primitive.

We exclude mechanisms that discourage ordering manipulation without making unfair orders invalid under the protocol. Examples are MEV mitigations at the application layer, mempool policies that proposers can ignore, and fee market rules or cryptographic incentives that only raise the cost of manipulation. Such soft mitigations currently dominate in practice (e.g., MEV-Boost~\cite{mev_boost}), but they provide no deterministic guarantee at the consensus layer. We focus on hard ordering rules that give verifiable, cryptographic bounds on adversarial reordering. We also exclude probabilistic mechanisms that accept any proposer order as valid.

The comparison tables (Tables~\ref{tab:FIFO_comparison} and~\ref{tab:random_comparison}) include only protocols that state an explicit fairness guarantee and specify their mechanism precisely enough to derive complexity. Rashnu~\cite{data-dependent_ordering} restricts fairness to data dependent transactions, SpeedyFair~\cite{speedyfair} speeds up an existing notion by decoupling ordering from consensus, and Travelers~\cite{timestamping_travelers} reduces dissemination cost. None proposes a new ordering rule, so we cite but do not tabulate them.

\subsection{Problem Setting}\label{sec:problem_setting}

We consider a setting in which nodes have equal opportunity to receive messages and participate in ordering, assuming no geographic, topological, or infrastructural advantage among nodes unless stated otherwise.

We separate economic motivation from the formal adversary model. MEV explains why rational validators may manipulate ordering, but our analysis assumes Byzantine behavior, which subsumes rational manipulation. We ask which guarantees survive when up to $f$ participants deviate arbitrarily. We prefer this to a game theoretic model for two reasons. First, Byzantine guarantees hold regardless of economic incentives. Incentives shift with market conditions and protocol upgrades, so guarantees that depend on an equilibrium are fragile. Second, MEV infrastructure such as MEV-Boost~\cite{mev_boost} creates complex games among many parties, and their equilibria are poorly understood. 

This model suffices for safety and liveness, but fair ordering needs a finer view of how Byzantine behavior affects ordering. We characterize the adversary through three capabilities.

\begin{enumerate}
    \item {\em Observation falsification}. A Byzantine node reports incorrect local observations that are used to order messages, such as fabricated timestamps, swapped arrival orders, or different reports to different nodes.
    \item {\em Scheduling}. Within the assumed network model (e.g., partial synchrony), the adversary influences delivery timing so that honest nodes observe different relative orders. Beyond that model, scheduling reduces to suppression.
    \item {\em Suppression}. A node withholds messages so that they are not disseminated before ordering. Suppression is usually treated separately, so we defer the details to Section~\ref{sec:suppression} and restrict our analysis to messages that honest nodes receive.
\end{enumerate}

These capabilities are not mutually exclusive. Together they capture how an adversary biases relative order without breaking consensus safety. We analyze each ordering rule by the capabilities it tolerates, and Table~\ref{tab:adv_capability} summarizes the result for each family.

\begin{table}[t]
\caption{Adversary Capability Tolerance by Protocol Family}
\label{tab:adv_capability}
\centering
\begin{threeparttable}
\renewcommand{\arraystretch}{1.2}
\begin{tabularx}{\textwidth}{@{}l *{3}{>{\centering\arraybackslash}X}@{}}
\toprule
Family & Observation falsification & Suppression & Scheduling \\
\midrule
FIFO voting & Partial ($\gamma$-majority) & Not addressed & Vulnerable \\
FIFO timestamping & Partial ($f$-robust) & Not addressed & Tolerant (clocks) \\
Random ordering & Addressed & Not addressed & Addressed \\
Blind ordering & Partial (content only) & Not addressed & Partial (non content) \\
\bottomrule
\end{tabularx}
\begin{tablenotes}[para,flushleft]
\footnotesize
\item \textbf{Note:} ``Partial'' = tolerated under stated conditions. ``Vulnerable'' = scheduling can decorrelate honest observations and undermine the guarantee. ``Tolerant (clocks)'' = tolerant given synchronized clocks and bounded network delay. For blind ordering, ``Partial'' means that only manipulation based on content is neutralized (Section~\ref{sec:blind_ordering}).
\end{tablenotes}
\end{threeparttable}
\end{table}

\section{Message Ordering Rules and Fairness Notions}
This section formalizes message ordering rules in the Byzantine fault tolerant setting and discusses what each rule requires and where it falls short.

\subsection{Defining Fair Ordering Rules}

Let an \emph{input set} $S$ be the set of messages to be ordered in one consensus instance, and \emph{metadata} $D$ be the information used by an ordering rule (local rankings, timestamps, or randomness). A deterministic \emph{message ordering rule} is written as $r(S,D)=O$, where $O$ is a total order over the messages in $S$. For FIFO via voting, $D$ is a collection of local arrival orders. For FIFO via timestamping, $D$ is a collection of timestamps or delay estimates. For random ordering, $D$ is the agreed randomness. We do not model blind ordering as a standalone rule $r$, because encryption only restricts what nodes can see.

Proposals that promote fairness probabilistically without applying an ordering rule~\cite{cohen2025fair,zhao2025mitigating,yahyaoui2025mitigating} are out of scope. Table~\ref{tab:taxonomy} summarizes the taxonomy. The ``central limitation'' column is our own classification, derived from the adversary capabilities and from the assumptions stated in the cited proposals.

\begin{table}[t]
\caption{Consensus-Layer Fair-Ordering Taxonomy Used in This Paper}
\label{tab:taxonomy}
\centering
\scriptsize
\setlength{\tabcolsep}{2.5pt}
\renewcommand{\arraystretch}{1.0}
\resizebox{\textwidth}{!}{%
\begin{tabular}{@{}p{2.4cm}p{3.0cm}p{2.8cm}p{3.7cm}p{3.7cm}@{}}
\toprule
Family & Ordering objective & Metadata used by the rule & Representative proposals & Central limitation \\
\midrule
FIFO via voting &
Preserve local arrival precedence when sufficiently many nodes agree &
Local rankings or pairwise precedence reports &
Aequitas~\cite{voting_ordering}, Quick~\cite{differential_order_fairness}, Themis~\cite{Themis_ordering}, streaming social choice~\cite{condorcet_ranked_pair} &
Scheduling can decorrelate honest local orders and create Condorcet cycles. \\
FIFO via timestamping &
Order by an agreed timestamp estimate for each message &
Local timestamps, inferred delays, or median timestamps &
Pomp\=e~\cite{timestamping1}, Wendy~\cite{timestamping_wendy}, Hashgraph~\cite{hashgraph}, Chronos~\cite{zhang2024chronos} &
Requires clock or delay assumptions; overlapping timestamp distributions can be manipulated. \\
Random ordering &
Give no included message systematic priority over another &
Agreed input set and public or generated randomness &
Helix~\cite{blind_fairness3_Helix,blind_fairness4_Helix}, $\Pi^3$~\cite{alpos2023eating}, BlindPerm~\cite{kavousi2023blindperm}, MEVade~\cite{piet2023mevade} &
Guarantee is conditional on inclusion; it does not force a proposer to include pending messages. \\
\bottomrule
\end{tabular}%
}
\end{table}

\subsection{FIFO Ordering Rule}
The most studied ordering rule is FIFO ordering~\cite{voting_ordering,age-aware_fairness,timestamp_reputation,voting_ordering_permissionless,differential_order_fairness,bounded_unfairness_ordering,Themis_ordering,anchor_based_efficient_ordering,data-dependent_ordering,condorcet_attack,condorcet_ranked_pair,timestamping1,timestamping_wendy,timestamping_wendy_extended,hashgraph,timestamping_travelers,median_sequence,efficient_timestamping,timestamping_and_TEE,blind_fairness9_and_delay_inference,decentralized_clock_network,kang2025fairdag}, which commits messages in the order in which the system first receives them.

As Kelkar et al.~\cite{voting_ordering} note, the time at which a message is {\em sent} cannot be measured without extra assumptions about the sender's network or a trusted timestamping service. FIFO ordering therefore targets the time at which messages are first {\em received}. 

FIFO ordering needs a well-connected peer-to-peer network to resist {\em scheduling} (Appendix~\ref{sec:roadmap}). If Byzantine nodes control the only links between groups of honest nodes, they can shape arrival orders without blocking delivery. No FIFO method based on local observations can then guarantee a stable order. Permissionless networks that are vulnerable to Eclipse and Sybil routing attacks easily violate this assumption, so FIFO via voting may be secure only in permissioned or tightly regulated networks. Two main methods achieve FIFO ordering, ranked voting and median timestamping.

\subsubsection{FIFO via Ranked Voting}
Agreeing on an order from different local preferences can be modeled as ranked choice voting. Several proposals collect local arrival rankings from each node and aggregate them into a message order~\cite{voting_ordering,voting_ordering_permissionless,differential_order_fairness,bounded_unfairness_ordering,Themis_ordering,anchor_based_efficient_ordering,data-dependent_ordering,cachin2024quick}. Unlike standard ranked choice voting, these algorithms must handle incomplete rankings, because not every node receives every message.

Given an unordered message set $S$, let $O^S_i$ be the ordered list of messages based on the arrival order of messages locally observed by node $i$. FIFO order via voting is defined as the following ordering rule:
\[FIFO(S, [O^S_1, O^S_2, ..., O^S_n])\]

\begin{figure}
    \centering
    \includegraphics[width=0.48\textwidth]{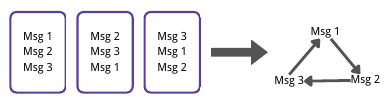}
    \caption{A Condorcet cycle from ranked voting: two of three nodes receive message 1 before 2, two receive 2 before 3, and two receive 3 before 1.}
    \label{fig:condorcet_cycle}
\end{figure}

Each node's local order acts as a ballot, and the majority preference decides the final order. Under an honest majority, aggregation tolerates a bounded amount of observation falsification. However, Condorcet cycles can arise~\cite{condorcets_paradox,voting_ordering}. Figure~\ref{fig:condorcet_cycle} shows three messages whose order cannot be decided because the vote is a draw.

Relaxed fairness notions have been proposed to tolerate Condorcet cycles. Batch order fairness~\cite{voting_ordering} and differential order fairness~\cite{differential_order_fairness} allow several messages to be treated as simultaneous, so that any order among them is acceptable.

\begin{definition}[$\gamma$-Batch Order Fairness] \label{def: batch order fairness}
Let $S$ be the input set ordered by a consensus instance, and let $m_1,m_2 \in S$ be two messages that were received by all honest nodes before the ordering decision for $S$. We say that a protocol satisfies \emph{$\gamma$-batch order fairness} for some $\gamma \in (0.5, 1]$ if the following holds. If at least a $\gamma$-fraction of all $n$ nodes received $m_1$ before $m_2$, then all honest nodes output $m_1$ no later than $m_2$.
\end{definition}

Following Aequitas~\cite{voting_ordering} and Themis~\cite{Themis_ordering}, $\gamma$ is measured over all $n$ nodes. The premise refers to actual reception, which Byzantine nodes cannot alter. They can only misreport it, and the resilience bounds in Table~\ref{tab:FIFO_comparison} account for such false reports.

Condorcet cycles can be arbitrarily long, so committing all members of a cycle at once harms liveness~\cite{voting_ordering}. One fix assumes an external synchrony bound~\cite{voting_ordering}, so that messages received far enough apart cannot belong to the same cycle. This costs latency and adds an assumption. Better approaches break cycles without violating FIFO fairness. Ranked pairs~\cite{ranked_pairs,condorcet_attack,condorcet_ranked_pair} builds a graph of pairwise preferences ordered by the strength of the majority and discards the edges that would create a cycle. The result is a deterministic total order without batching. Themis~\cite{Themis_ordering} unspools batches instead. It commits the messages that do not belong to a cycle and defers the rest, which preserves liveness and leaves the leader less room to reorder.

Ramseyer and Goel~\cite{condorcet_ranked_pair} strengthen this line of work in two ways. Batch order fairness gets easier to satisfy as batches grow, and one giant batch satisfies it vacuously, so they require batches to be as small as the votes allow. They also treat local orders as unbounded streams and adapt ranked pairs to extend the committed sequence incrementally, with liveness guaranteed, on top of the dissemination protocols of prior work.

Cycles also threaten fairness. Treating a batch as simultaneous is not enough for blockchains, which require a total order. Unless the order within a batch is fixed by a deterministic and verifiable tie breaking rule, a proposer can arrange the batch to its own advantage.

Consider three messages $a,b,c$, seven nodes, and one Byzantine node. A scheduling adversary can delay dissemination so that the honest local orders split, for example into $a\!\prec\!b\!\prec\!c$, $b\!\prec\!c\!\prec\!a$, and $c\!\prec\!a\!\prec\!b$. The Byzantine node then reports whichever ranking strengthens the cycle. The result is a majority for each of $a\!\prec\!b$, $b\!\prec\!c$, and $c\!\prec\!a$. If $\{a,b,c\}$ is placed in one batch, the executed order is decided by the tie breaking rule. A leader can still profit from where its own transaction lands relative to the victim's. This is the {\em scheduling} capability at work~\cite{condorcet_attack,park2025frontrunning}. A complementary direction prevents cycles from forming in the first place, for example through more frequent propagation~\cite{condorcet_attack}.

In terms of the adversary capabilities, FIFO via voting tolerates bounded observation falsification through aggregation. It remains vulnerable to suppression, when a message reaches too few honest nodes, and to scheduling, when adversarial timing makes honest rankings disagree even though every message is eventually delivered. It also lacks public verifiability. Local arrival orders are subjective, so an external observer cannot tell honest metadata from fabricated reports.

\subsubsection{FIFO via Timestamping}
Another FIFO approach assigns each message a timestamp and orders by it~\cite{timestamping1,timestamping_wendy,timestamping_wendy_extended,hashgraph,timestamping_travelers,median_sequence,efficient_timestamping,timestamping_and_TEE,blind_fairness9_and_delay_inference,decentralized_clock_network,zhang2024chronos,segalinifair}. Existing proposals timestamp in two ways.

\paragraph{Inference Timestamping}
The first infers a message's arrival time from known network delays between peers~\cite{timestamping_and_TEE,blind_fairness9_and_delay_inference}, assuming that these delays are relatively stable.

Both proposals~\cite{timestamping_and_TEE,blind_fairness9_and_delay_inference} assume that each message is broadcast to all nodes. When delays are unstable, timestamps fall outside the expected range, nodes cannot agree, and liveness is lost. Given these strict conditions, this category provides no fairness guarantee.

\paragraph{Median Timestamping}
The second uses the median of the nodes' local reception timestamps~\cite{timestamping1,timestamping_wendy,timestamping_wendy_extended,hashgraph,median_sequence,efficient_timestamping,decentralized_clock_network}. The median limits the effect of malicious timestamps ({\em observation falsification}), because it always lies within the range of honest reception times.

Given a set of messages $S$, let $T^S_i$ denote the map of timestamps at which every message of $S$ is locally observed by node $i$. FIFO ordering via median timestamping is defined as the following ordering rule:
\[FIFO(S, [T^S_1, T^S_2, ..., T^S_n])\]

The median comes from a single node (or two). In Table~\ref{tab:median_timestamping}, message~1 precedes message~2 by median. A malicious node~2 can reverse this by reporting 3:03 for message~1 and 3:02 for message~2. This is believed to be hard in practice~\cite{timestamping1,timestamping_wendy}, since the attacker must hold exactly the median position, but it cannot be ruled out.

\begin{table}
    \centering
    \caption{An Example of Taking the Median Timestamp to Order Messages}
    \begin{tabular}{c|ccc|c}
        Message & Node 1 & Node 2 & Node 3 & Output\\ \hline
        Msg 1 & 3:00 & 3:02 & 3:04 & 3:02 \\
        Msg 2 & 3:01 & 3:03 & 3:05 & 3:03 \\
    \end{tabular}
    \label{tab:median_timestamping}
\end{table}

Existing definitions~\cite{timestamping_wendy,timestamping1,Themis_ordering} of this guarantee do not cover this case. Kelkar et al.~\cite{Themis_ordering} show that if all honest nodes receive $m_1$ before any honest node receives $m_2$, then all honest nodes output $m_1$ before $m_2$. They do not consider that an adversary can swap the two median timestamps. We therefore give a definition that captures this case.

\begin{definition}[$f$-Robust Median Order] \label{def: robust median order}
Let $m_1,m_2 \in S$ be two messages that are both received by all honest nodes before the ordering decision for $S$. Let $n=3f+1$ and let the reported median use the lower-median rank $k=\lceil n/2\rceil$ over the $n$ reported timestamps. For $j \in \{1,2\}$, let $T^H_{(q)}(j)$ denote the $q$-th smallest honest timestamp for $m_j$.

We say that the relative order $m_1 \prec m_2$ is \emph{$f$-robust under median timestamping} if:
\[
T^H_{(k)}(1) < T^H_{(k-f)}(2).
\]
\end{definition}
This condition is sufficient against an adversary controlling up to $f$ timestamp reports per transaction. To make $m_1$ appear later, Byzantine nodes can report arbitrarily large timestamps for $m_1$, in which case the reported median is no larger than the $k$-th honest timestamp of $m_1$. To make $m_2$ appear earlier, they can report arbitrarily small timestamps for $m_2$, in which case the reported median is no smaller than the $(k-f)$-th honest timestamp of $m_2$. The inequality is therefore a sufficient condition under which Byzantine reports cannot reverse the median order. It is stated over the honest timestamps, which are not known at run time, so it serves only as an analysis tool. If the honest timestamps of $m_1$ and $m_2$ overlap substantially, the condition fails, and the median rule no longer supports a robust ordering claim. In practice, global latency and jitter make the honest timestamps of concurrently broadcast transactions overlap. This strict separation is therefore hard to satisfy without synchronized clocks and very stable delays. Appendix~\ref{sec:median_example} gives a numerical example of the definition.

Comparing timestamps across nodes is meaningless unless local clocks are synchronized, so all median timestamping proposals require clock synchronization. This is feasible with BFT clock synchronization protocols~\cite{bft_clock_sync,evangelos2020proof}. Byzantine agreement on a median has been studied~\cite{median_agreement} and applied to this setting~\cite{decentralized_clock_network}. The real goal is different, however. Ideally the agreed value would be the smallest honest timestamp, which is the closest to the actual send time. How to minimize the gap between the actual and the derived timestamp remains an open problem.

In terms of the adversary capabilities, FIFO via timestamping partially tolerates observation falsification through the median, but only when reports are authenticated and the timestamp distributions are well separated. It remains vulnerable to suppression, when too few honest timestamps exist, and to scheduling, when adversarial delays shift the distributions within the network model. Like voting, it lacks public verifiability, since an external party cannot check whether the reported timestamps are genuine.

\subsubsection{Comparison of FIFO Protocols}
Table~\ref{tab:FIFO_comparison} (Appendix~\ref{sec:detailed_comparison}) compares FIFO protocols by design, assumptions, and complexity. 

\subsection{Random Ordering Rule}

\subsubsection{Methods for Random Ordering}
Another approach to fair ordering is to order messages randomly~\cite{blind_fairness3_Helix,blind_fairness4_Helix,random_fairness1,helix_with_reputation,alpos2023eating,kavousi2023blindperm,piet2023mevade}. 

Given a set of messages $S$, denote $R$ as the randomness agreed upon by all nodes. Random ordering is defined as:
\[Random(S, R)\]

The core requirement is randomness that all nodes agree on. Existing proposals~\cite{blind_fairness3_Helix,blind_fairness4_Helix,random_fairness1,helix_with_reputation,alpos2023eating,kavousi2023blindperm,piet2023mevade} use information from the previous block as public randomness, hash it with each message to obtain a pseudorandom value, and place the lowest values first.

Early proposals~\cite{blind_fairness3_Helix,blind_fairness4_Helix,random_fairness1,helix_with_reputation} are not deterministic and require no common input set $S$. A proposer orders its own pool, and peers verify the result by estimating how likely it is under honest behavior. The agreed order may still contain messages whose hashes were crafted to land in specific positions, so no deterministic guarantee follows.

Recent work instead permutes the committed set of messages before execution~\cite{alpos2023eating,kavousi2023blindperm,piet2023mevade}, which enforces randomness deterministically. The guarantee is \emph{symmetric ordering conditional on inclusion}. Once $S$ is fixed, no included message gets systematic priority over another. This differs from inclusion fairness, which would require pending messages to be in $S$ in the first place. 

\begin{definition}[Permuted Ordering Conditional on Inclusion] \label{def: permuted ordering}
Let $M = \{m_1, m_2, \dots, m_n\}$ be a set of $n$ committed messages, and let $\sigma: M \to \{1, \dots, n\}$ be a random bijection drawn uniformly at random from the set of all permutations over $M$. The protocol enforces a \emph{permuted ordering} if for all distinct messages $a, b \in M$, the following holds:
\[
\Pr[\sigma(a) < \sigma(b)] = \Pr[\sigma(b) < \sigma(a)] = \frac{1}{2}.
\]
\end{definition}

This condition is the main limitation of random ordering as an MEV defense. Protocols permute within a block but rarely constrain which messages enter it. A proposer can still front-run by including its own transaction and excluding the victim's. Control over inclusion also enables censorship, delay, and MEV across several blocks. Random ordering thus provides symmetry but not inclusion fairness, and offers almost no protection against a proposer that controls $S$.

In terms of the adversary capabilities, random ordering is immune to falsified arrival observations and less sensitive to scheduling, because arrival metadata plays no role once $S$ is fixed. Its main weakness is suppression. Whoever controls $S$ can still exclude, delay, or selectively include messages before the permutation. Unlike FIFO, random ordering based on VRFs or VDFs produces auditable proofs that the permutation followed the agreed randomness. This gives public verifiability of the order, though not of inclusion.

\subsubsection{Comparison of Random Ordering Protocols}
Table~\ref{tab:random_comparison} (Appendix~\ref{sec:detailed_comparison}) compares random ordering proposals. The permutation approach has the lowest ordering complexity, needs no synchronized clocks, and integrates with various consensus protocols~\cite{alpos2023eating,piet2023mevade}. It is a lightweight defense against manipulation of the order after inclusion, but not against manipulation of inclusion itself.

\subsection{Comparison of Fairness Notions}
The fairness notions in this paper pursue different objectives. FIFO guarantees and permuted ordering are mutually exclusive. Within FIFO, $\gamma$-batch order fairness and $f$-robust median order are closely related but logically independent. Both preserve relative order under partial adversarial influence, but they rely on different conditions. $\gamma$-batch fairness is based on majorities. It preserves order when a sufficient fraction of nodes agree on precedence. $f$-robust median order is based on separation. It preserves order when the timestamp distributions are far enough apart to withstand $f$ falsified reports. Under strong synchrony, agreement among nodes correlates with separation, but neither implies the other.

Moving from $\gamma$-batch to $f$-robust median order replaces majority reasoning with robustness to bounded falsification, at the cost of clock and timing assumptions. Permuted ordering gives up arrival order for symmetry. It drops arrival metadata and the requirement of a well-connected network, but it no longer preserves order. The three form no hierarchy but reflect different priorities under different assumptions.

To see that the two notions are logically independent, consider $n=13$ ($f=2$) and $\gamma=0.8$, which satisfies the voting resilience bound $n \geq \frac{2f(\gamma+1)}{2\gamma-1} = 12$. The $\gamma$-batch premise requires $\lceil 0.8 \cdot 13\rceil = 11$ of the $13$ nodes to receive $m_1$ first, for instance all $11$ honest nodes. In one direction, all honest nodes may receive $m_1$ first while their timestamps for the two messages interleave within $1$ to $2$~ms. The separation condition $T^H_{(7)}(1) < T^H_{(5)}(2)$ then fails, and two Byzantine reports can reverse the median order. In the other direction, scheduling can make five honest nodes receive $m_2$ first while the separation condition still holds. Then at most $8 < 11$ nodes received $m_1$ first, so $\gamma$-batch fairness imposes no constraint, while the median order remains $f$-robust.

\section{Blind Ordering as an Orthogonal Restriction}\label{sec:blind_ordering}

Blind ordering proposals~\cite{blind_fairness1,blind_fairness2,blind_fairness3_Helix,blind_fairness4_Helix,blind_fairness5,blind_fairness6,blind_fairness7,blind_fairness8,blind_fairness9_and_delay_inference,mempool_privacy,alea-bft,ciampi2024universal,ciampi2025universally,zhang2022flash,bormet2024beat,choudhuri2024practical,agarwal2024efficiently,misra2024towards,rivaseahorse,bormet2024beat,choudhuri2024practical,agarwal2024efficiently} use threshold cryptography~\cite{secret_sharing,threshold_cryptosystems} to encrypt messages before they are committed. Only a threshold of nodes can decrypt them afterward, so the content stays hidden during ordering. Time lock puzzles~\cite{khajehpour2023mitigating} are an alternative, but their solving time is unstable~\cite{li2024blockchain}, which risks revealing content too early.

Blind ordering is not an ordering rule on its own, since nodes can still order messages arbitrarily. It is a restriction, or privacy primitive, that is applied alongside random~\cite{blind_fairness3_Helix,blind_fairness4_Helix,random_fairness1,helix_with_reputation,kavousi2023blindperm,piet2023mevade} or FIFO~\cite{blind_fairness9_and_delay_inference,decentralized_clock_network,timestamping_travelers} ordering.

Blind ordering promotes fairness because a proposer cannot censor or reorder messages based on their content before decryption. It does not ensure censorship resistance, however. A proposer can still exclude encrypted messages based on sender identity, fee, size, timing, or network origin. Three properties should be distinguished. \emph{Content privacy} hides what a transaction does until it is committed. \emph{Inclusion fairness} constrains which pending messages must be included in $S$. \emph{Ordering fairness} constrains the sequence $O$. Blind ordering provides content privacy and supports ordering fairness when combined with FIFO or random ordering, but it does not provide inclusion fairness.

In terms of the adversary capabilities, blind ordering removes the value of observing content, which weakens manipulation motivated by MEV before decryption. It does not prevent falsified metadata in FIFO or timestamp rules unless those rules authenticate their metadata. Nor does it prevent suppression or scheduling based on features other than content, such as sender, fee, timing, size, or ciphertext hash. Threshold encryption does give publicly auditable proof that content was hidden during ordering, a form of public verifiability that FIFO approaches lack.

A further weakness is the window between decryption and the next round. Once transactions are revealed, the builder of the next round can see their contents and place its own transactions accordingly in later blocks. Countering this form of MEV across blocks requires either encryption that spans rounds (rolling threshold encryption) or a deterministic rule applied after decryption that removes the proposer's discretion.

\section{A Common Integration Pattern for Embedding Fairness into Consensus Protocols}

Several fair ordering protocols~\cite{timestamping_wendy,bounded_unfairness_ordering,alpos2023eating,piet2023mevade} share a common design pattern. Fairness is added to an existing consensus protocol as a modular refinement rather than built into a completely new protocol. This section describes that pattern as a \emph{reusable design framework}. It identifies the components a protocol must implement and states the conditions under which safety and liveness are preserved (Propositions~\ref{prop:safety} and~\ref{prop:liveness}, formalized in Appendix~\ref{sec: safety and liveness}). Appendix~\ref{sec:leaderless_integration} shows how the pattern applies to leader based and leaderless protocols. We do not claim a black box transformation theorem. The guarantees depend on the validation, synchronization, and recovery conditions stated below.

\subsection{Adding Order Fairness to Consensus Protocol}

Let $\mathcal{P}$ be the class of BFT consensus protocols that satisfy standard safety and liveness but no order fairness property (Definitions~\ref{def: batch order fairness},~\ref{def: robust median order},~\ref{def: permuted ordering}). Each $P \in \mathcal{P}$ assumes a set of nodes, a dissemination layer, and a decision procedure producing a committed transaction sequence.

To describe the pattern, consider augmenting a protocol $P$ with a deterministic fair ordering rule $r$. The resulting protocol, denoted $P_{\text{fair}}$, derives candidate transaction orders by applying $r$ to a synchronized input set and metadata. The augmentation has two components:

\paragraph{Component 1: State Synchronization}
Nodes may hold inconsistent views of the pending messages because of asynchronous delivery, selective dissemination, or adversarial interference. A synchronization step therefore establishes a common view of the candidate set $S$ and the metadata $D$ before ordering, either through explicit communication or through dissemination during the consensus rounds. Observation falsification is tolerated only insofar as the chosen rule tolerates it.

The cost of this step depends on the synchronization architecture, and two patterns dominate. In all-to-all designs (e.g., Aequitas~\cite{voting_ordering}), every node disseminates its complete local ordering to all peers with reliable broadcast guarantees, requiring $O(n^3)$ communication. This cost dominates latency and is hard to scale beyond small committees. In leader mediated designs (e.g., Themis~\cite{Themis_ordering}), replicas send their signed local orderings to the current leader. The leader proposes an order together with the supporting metadata, and replicas validate the proposal by recomputing the ordering rule. This requires $O(n^2)$ communication in the common case. Primitives akin to Byzantine Reliable Broadcast are needed only on the fallback path that recovers from a leader that equivocates on metadata. Synchronization can also be amortized by embedding metadata exchange into leader election or the finalization messages of the previous round (Appendix~\ref{sec:leaderless_integration}). 

\paragraph{Component 2: Local Rule Execution}
Once $S$ and $D$ are synchronized, each honest node applies the deterministic rule $r$ on its own to produce the ordered output (Figure~\ref{fig:fair_consensus_overview}). Because $r$ is deterministic and the inputs are agreed, all honest nodes obtain the same result without a central proposal or aggregation phase. A leader therefore cannot impose an arbitrary order over the synchronized input. The limitations regarding suppression and scheduling remain. In short, the integration changes how $P$ derives a proposal but not how $P$ agrees on it.

\begin{figure}
    \centering
    \resizebox{0.82\textwidth}{!}{%
    \begin{tikzpicture}[
      box/.style={draw=violet!75!black, very thick, rounded corners, align=center,
                  inner sep=3pt, minimum height=8.5mm, minimum width=26mm},
      bld/.style={box, font=\bfseries},
      dsh/.style={box, dashed, font=\bfseries},
      lbl/.style={font=\footnotesize, fill=white, inner sep=1pt},
      ar/.style={-{Stealth[length=2.4mm]}, very thick, violet!75!black}]
      \node[box] (root) {How to order messages};
      \node[box] (fifo) at ($(root)+(-3.0,-1.85)$) {How to FIFO};
      \node[bld] (rand) at ($(root)+(3.2,-1.85)$) {Random ordering};
      \node[bld] (vote) at ($(fifo)+(-1.6,-1.85)$) {FIFO through\\Ranked Voting};
      \node[bld] (ts)   at ($(fifo)+(1.9,-1.85)$) {FIFO through median\\timestamping};
      \node[dsh] (blind) at ($(rand)+(0,-1.85)$) {Optional add-on:\\blind ordering};
      \draw[ar] (root) -- (fifo) node[lbl, pos=0.5]{FIFO};
      \draw[ar] (root) -- (rand) node[lbl, pos=0.5]{Randomly};
      \draw[ar] (fifo) -- (vote) node[lbl, pos=0.5]{voting};
      \draw[ar] (fifo) -- (ts)   node[lbl, pos=0.5]{timestamping};
    \end{tikzpicture}}
    \caption{Overview of the ordering mechanisms and the optional blind ordering restriction systematized in this paper.}
    \label{fig:fair_consensus_overview}
\end{figure}

\subsection{Quantitative Overhead Estimates}\label{sec:overhead}

We further demonstrate the concrete cost by estimating the communication and ordering overhead of each family at two committee sizes, a small one ($n=10$, $m=100$) and a large one ($n=100$, $m=1000$). We assume BLS signatures (96\,B), SHA-256 hashes (32\,B), and 256\,B transactions, and report costs relative to vanilla HotStuff~\cite{hotstuff} without fair ordering. Table~\ref{tab:overhead} (Appendix~\ref{sec:detailed_comparison}) reports the estimates, and Appendix~\ref{sec:overhead_model} gives the cost model.

Voting based protocols (Aequitas, Themis) have the highest overhead at scale. At $n=100$ and $m=1000$, Aequitas needs over $440$\,MB per round ($18\times$ the baseline) and $10^8$ pairwise comparisons. Voting based FIFO is therefore impractical for large committees without batching or compression. Timestamping protocols (Pomp\=e, Wendy) are far cheaper in communication, at about $1.7\times$ the baseline for both sizes, because timestamps are small (8\,B) and ordering reduces to a sort. Their median computation per transaction does grow to $109\times$ at $n=100$. Hashgraph also keeps communication moderate, at about $3.5\times$, because dissemination is folded into its gossip layer. Random ordering (Helix) inverts the tradeoff. Its communication is high ($13.7\times$, due to encrypted broadcast and all-to-all verification), but its ordering cost is a minimal $O(m\log m)$.

All protocols we analyze incur at least $1.5\times$ the baseline communication. Permutation modules such as $\Pi^3$~\cite{alpos2023eating} and MEVade~\cite{piet2023mevade} reuse existing consensus messages and add little communication, at the price of offering only symmetry conditional on inclusion. The table also compares systems of different kinds under one per round metric, namely full protocols, a pre-protocol (Wendy), a gossip system (Hashgraph), and a permutation module (BlindPerm). Fair ordering nevertheless carries a real communication cost, which motivates the directions in Appendix~\ref{sec:roadmap}, namely committee subsampling, filtering of order insensitive transactions, and outsourced sequencing.

\subsection{Dissemination, Eligibility, and Censorship}\label{sec:suppression}

Whether a message is ordered fairly depends on three separate problems. The first is \emph{reliable dissemination}, a property of the network. After GST, a message received by one honest node must reach all honest nodes within bounded time. If an adversary can keep a message below the dissemination threshold indefinitely, ordinary liveness already fails, and no ordering mechanism can include data that honest nodes do not have. Protocols discharge this assumption by having clients send to all nodes~\cite{Themis_ordering,differential_order_fairness} or by assuming bounded propagation delay~\cite{voting_ordering}.

The second is \emph{eligibility}, which is part of the ordering pipeline. A message enters $S$ only once enough nodes report it, at least $f+1$ honest ones, so that Byzantine nodes can neither suppress it nor control its metadata. Themis, for example, orders a transaction once $n-f$ nodes report it and defers it otherwise. A message below the threshold is deferred.

The third is \emph{censorship of eligible messages}. Ordering rules constrain only the sequence over $S$, so a proposer can omit a victim's transaction or propose an empty block to delay it. Inclusion fairness mechanisms, such as Ethereum's inclusion lists~\cite{blockchain_censorship} and threshold encrypted mempools that force ciphertext inclusion~\cite{blind_fairness1,blind_fairness2}, address this third problem only. They are beyond our scope, but any deployed fair ordering system needs them for end-to-end MEV protection. Blind ordering helps partially, since executing $r$ without seeing content removes the incentive to omit messages based on content.

Finally, recent proposals enforce ordering from a single node without all-to-all synchronization via trusted execution environments (TEEs)~\cite{blind_fairness6,TEE_mempool1,TEE_mempool2} or zero-knowledge proofs~\cite{Themis_ordering}, but these add trust or computational costs (ZK ordering requires expensive proof generation with no demonstrated low latency evaluation) that may conflict with decentralization or latency goals. Our framework therefore prioritizes all-to-all synchronization and deterministic execution.

\subsection{Mapping Fair Ordering to Proposer-Builder Separation}\label{sec:pbs}
The integration pattern relies on a committee of honest nodes to collect metadata and execute $r$. This clashes with Proposer-Builder Separation (PBS), which dominates in Ethereum and outsources block construction to centralized builders. Builders operate outside consensus and maximize MEV. They cannot be trusted to report arrival times honestly or to construct $S$ without omissions. FIFO rules are therefore incompatible with unregulated builders.

Enforcing fair ordering under PBS requires either taking ordering out of the builder's hands or forcing compliance by technical means. Builders could be required to construct blocks inside trusted execution environments, as in the SUAVE architecture~\cite{TEE_mempool1}, or to sequence ciphertexts blindly, which is subject to the decryption window discussed in Section~\ref{sec:blind_ordering}. Neither path has been deployed at production scale for fair ordering, so fair ordering protocols remain largely theoretical in PBS environments.

\subsection{Practitioner Guidance}\label{sec:guidance}

Combining the overhead estimates (Table~\ref{tab:overhead}) with the adversary capability analysis (Table~\ref{tab:adv_capability}), we map five deployment scenarios to recommended approaches. Table~\ref{tab:guidance} (Appendix~\ref{sec:detailed_comparison}) gives the full rationale. A small permissioned committee ($n\le20$) that must preserve arrival order is best served by FIFO via voting (Themis, about $1.5\times$ the baseline). A committee of any size that needs FIFO at low communication cost should use timestamping (Pomp\=e or Wendy, about $1.7\times$), provided synchronized clocks are available. A large committee ($n>50$) that does not need arrival order should use permutation based random ordering, which costs $O(m)$ and needs no clocks, but is conditional on inclusion. A permissionless setting that minimizes trust and needs public verifiability should combine blind ordering with random permutation. PBS and other architectures with centralized builders require blind ordering or builders constrained by TEEs, since standard fair ordering is incompatible with unregulated builders (Section~\ref{sec:pbs}). All recommendations assume $n\ge3f+1$ (voting needs $n\ge4f+1$ at $\gamma=1$). None provides inclusion fairness, which needs complementary mechanisms such as inclusion lists.

\section{Conclusion}
This paper systematizes mechanisms at the consensus layer for fair message ordering in distributed ledgers. We introduce an adversary capability framework that decomposes Byzantine behavior into observation falsification, suppression, and scheduling, and we analyze every ordering family against these capabilities. We provide a new formalization, the $f$-Robust Median Order, which identifies a gap in existing definitions of fairness based on timestamps. We also state explicit safety and liveness conditions for integrating fair ordering into BFT consensus. For the protocols and parameters we analyze, fair ordering costs $1.5\times$ to $18\times$ the communication of vanilla consensus, and ordering computation varies by up to five orders of magnitude across families. Together with the taxonomy and comparison tables, these results offer a structured reference for designing fair and efficient distributed ledgers.

\section*{Acknowledgments}
This project is funded by a UKRI Future Leaders Fellowship (MR-/W009560-/1): `\emph{Digitally Assisted Collective Governance of Smart City Commons--ARTIO}'.

\bibliographystyle{splncs04}
\bibliography{references}

\appendix
\section{Numerical Example for $f$-Robust Median Order}\label{sec:median_example}

To illustrate both Definition~\ref{def: robust median order} and its gap relative to prior work, consider a concrete example with $n=4$ ($f=1$). The median rank is $k=\lceil 4/2\rceil=2$, and $k-f=1$. Suppose three honest nodes receive $m_1$ at times $\{100, 105, 110\}$~ms and $m_2$ at $\{103, 108, 113\}$~ms. The condition requires $T^H_{(2)}(1) < T^H_{(1)}(2)$, i.e., $105 < 103$, which fails. The single Byzantine node can report $m_1$ at $200$~ms and $m_2$ at $0$~ms. With the lower median ($k=2$), the reported medians become $105$ for $m_1$ (the second smallest of $\{100,105,110,200\}$) and $103$ for $m_2$ (the second smallest of $\{0,103,108,113\}$), reversing the honest majority order. This attack succeeds even though all three honest nodes observed $m_1$ before $m_2$.

The condition of Kelkar et al.~\cite{Themis_ordering} instead requires all honest nodes to receive $m_1$ before any honest node receives $m_2$, i.e., $\max_i t_i(m_1) < \min_j t_j(m_2)$. Our example violates this precondition. The last honest receipt of $m_1$ is at $110$~ms, but the first honest receipt of $m_2$ is at $103$~ms. Their guarantee is therefore silent on this case. Yet every honest node individually observed $m_1$ before $m_2$ ($100 < 103$, $105 < 108$, $110 < 113$), so a practitioner might reasonably expect the order to hold. Definition~\ref{def: robust median order} fills this gap. It correctly flags the case as unsafe ($105 \not< 103$), because it accounts for the adversary's ability to shift a median by up to $f$ rank positions instead of relying on a global condition that the two sets of timestamps do not overlap.

The separation condition becomes hard to satisfy as $n$ grows. At $n=100$ ($f=33$) it requires $T^H_{(50)}(1) < T^H_{(17)}(2)$. In words, the 50th earliest honest receipt of $m_1$ must precede the 17th earliest honest receipt of $m_2$. Across a global network, where latency between continents is around 200~ms, two transactions broadcast within a few hundred milliseconds of each other will rarely meet this requirement. Roughly, their broadcasts must be separated by at least twice the maximum round trip time. The $f$-robust guarantee is therefore most meaningful for transactions that are not concurrent, which is also the regime where front-running is easiest to identify. For concurrent transactions with overlapping honest timestamps, the definition serves as a negative result. Median timestamping cannot provably prevent manipulation of their order, and complementary mechanisms such as blind ordering are needed.

\section{Integration with Leader Based and Leaderless Protocols}\label{sec:leaderless_integration}

\begin{figure}
    \centering
    \includegraphics[width=0.55\textwidth]{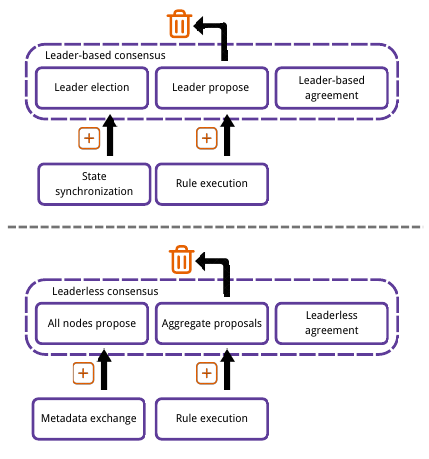}
    \caption{Adding fair ordering components to leader based (top) and leaderless (bottom) consensus}
    \label{fig:factory_process}
\end{figure}

Both leader based and leaderless consensus protocols fit naturally into this integration pattern. Figure~\ref{fig:factory_process} compares how the pattern applies to each protocol family.

\paragraph{Leader Based Consensus} \label{par: leader based consensus}

A leader based consensus protocol (e.g., HotStuff~\cite{hotstuff}, Alea~\cite{alea-bft}) proceeds through the following phases per round:

\begin{enumerate}
    \item Proposal: A designated leader collects a set of pending transactions $S$ from its local pool and produces an ordered sequence $O$ from an arbitrary ordering policy.
    \item Agreement: The leader broadcasts the proposal $O$ to other nodes. If they validate it, they vote to accept it. The protocol reaches consensus on $O$.
\end{enumerate}

After adding the fair ordering components, the process is as follows:

\begin{itemize}
    \item At the beginning of the consensus round (e.g., during leader election) or at the last phase of the last consensus round, all nodes exchange $S_i$ and $D_i$ to synchronize the input set $S$ and associated metadata $D$.
    \item Once $S$ and $D$ are agreed upon, each honest node independently computes $O = r(S, D)$ using the fair, deterministic ordering rule $r$.
    \item The result $O$ is passed into the standard voting and agreement phases of the protocol, unmodified.
\end{itemize}

In leader based consensus, state synchronization can often be folded into leader election or into the finalization messages of the previous round, which keeps the added latency small.

\paragraph{Leaderless Consensus}

Leaderless consensus protocols (e.g., Narwhal/Bullshark~\cite{sui_consensus}, Avalanche~\cite{avalance_consensus}, Mysticeti~\cite{sui_mysticeti}) typically follow a different structure:

\begin{enumerate}
    \item Exchange proposals: Each node $i$ proposes a local input set $S_i$ of messages.
    \item Aggregation: The protocol aggregates proposals $S_1, ..., S_n$ into a combined set $S$; applies a predefined ordering policy to derive a sequence $O$.
    \item Agreement: Nodes reach consensus on $O$ using the standard agreement procedure of the protocol.
\end{enumerate}

This process already meets the communication requirements of the integration pattern. Only the message exchange and the ordering rule change, as follows:

\begin{itemize}
    \item The exchange proposal phase is also used to exchange metadata $D_i$. As a result, honest nodes synchronize on a common input set $S$ and metadata $D$.
    \item The originally defined aggregation rule is replaced by the fair ordering rule $r$. Each node locally computes $O = r(S, D)$.
    \item The agreed upon output $O$ is then passed into the standard agreement procedure of the protocol.
\end{itemize}

\section{Safety and Liveness Conditions}\label{sec: safety and liveness}

The integration preserves agreement safety only when validation is deterministic and metadata is authenticated. Its liveness depends on further conditions introduced by synchronization and by executing the fair rule. We state these as explicit design guidelines with verifiable preconditions. 

\begin{proposition}[Safety Preservation]\label{prop:safety}
Let $P$ be a BFT consensus protocol with $n \geq 3f+1$ nodes (or $n \geq 4f+1$ for voting based FIFO) satisfying agreement safety (no two honest nodes commit different outputs). Let $P_{\text{fair}}$ be the augmented protocol obtained by the integration pattern above. If the following conditions hold:
\begin{enumerate}
    \item The ordering rule $r$ is a deterministic function: for any input $(S,D)$, all honest nodes compute the same output $O = r(S,D)$;
    \item Every proposed batch includes authenticated metadata $D$, and each honest node independently validates the batch by recomputing $r(S,D)$ before voting;
    \item The agreement mechanism of $P$ (e.g., quorum certificates, voting) is unmodified in $P_{\text{fair}}$;
    \item The synchronization phase (metadata collection) does not introduce new equivocation vectors beyond those already handled by $P$'s agreement protocol;
\end{enumerate}
then $P_{\text{fair}}$ satisfies agreement safety.
\end{proposition}

\noindent\emph{Argument.} Since $r$ is deterministic and every honest node validates by recomputing it, honest nodes vote only for batches that match their own $r(S,D)$. Identical $(S,D)$ yields identical $O$. Divergent $(S,D)$, whether due to partitions or Byzantine metadata, yields different $O$ and hence no shared vote. The quorum requirement of $P$ then prevents conflicting commits, and condition~(4) keeps synchronization within the failure model that $P$ already handles. Safety thus reduces to the safety of $P$, the determinism of $r$, and the containment of synchronization.

\begin{proposition}[Conditional Liveness]\label{prop:liveness}
Let $P$ be a BFT consensus protocol with $n \geq 3f+1$ nodes satisfying liveness (honest transactions are eventually committed after GST in the partial synchrony model). The augmented protocol $P_{\text{fair}}$ satisfies liveness if the following additional conditions hold:
\begin{enumerate}
    \item \textbf{Bounded metadata:} The metadata object $D$ has bounded size $|D| \leq B$ for a protocol-specified constant $B$;
    \item \textbf{Input stabilization:} After GST, honest nodes eventually construct a common candidate set $S$ and authenticated metadata $D$ within bounded rounds;
    \item \textbf{Deterministic tie breaking:} The ordering rule $r$ produces a unique total order for every valid $(S,D)$, including ties;
    \item \textbf{Bounded equivocation recovery:} If a Byzantine node sends inconsistent $(S,D)$ views to different honest nodes, equivocation detection and exclusion terminate within a bounded number of rounds $R_{\text{max}}$ after GST;
    \item \textbf{Non-abusive exclusion:} The exclusion mechanism targets only provably equivocating nodes (i.e., nodes for which conflicting signed messages exist), preventing weaponization against honest nodes that appear inconsistent due to pre-GST asynchrony.
\end{enumerate}
\end{proposition}

\noindent\emph{Argument.} Conditions (1) to (3) ensure that the fair ordering step produces a well defined proposal in bounded time. Condition (4) bounds recovery when equivocation stalls a round. Honest nodes detect inconsistent $(S,D)$ by exchanging hashes $h(S_i,D_i)$ and exclude the offender in the next round. Conditions (4) and (5) are in tension before GST, when delayed messages can make an honest node look as if it equivocated. Condition (5) resolves this by requiring cryptographic evidence, namely conflicting signed statements. After GST such evidence propagates in bounded time. Under all five conditions the liveness of $P$ is preserved, because the augmentation adds only a bounded number of rounds before a valid proposal is derived.

\section{Fair Ordering Research Roadmap}\label{sec:roadmap}
This section highlights open questions grounded in the analytical gaps identified above.

\subsection{Network Topology Requirements for FIFO}
Our analysis shows that FIFO ordering depends on the dissemination network, not only on the ordering rule or the consensus mechanism. Even when all honest nodes eventually receive the same messages, suppression and scheduling can shape their relative arrival order. Current proposals assume a well connected network with several independent paths, but rarely state or analyze this assumption formally. An important direction is to characterize the minimal topology and dissemination conditions under which FIFO guarantees are meaningful. Given a graph $G$, an adversarial set $B$, and a dissemination protocol $D$, which connectivity and relay conditions keep honest local orders correlated enough for $\gamma$-batch fairness to matter?

\subsection{Adding Flexibility to Order Insensitive Messages}
Existing proposals fairly order all messages regardless of order sensitivity. Distinguishing order sensitive messages could reduce latency and increase throughput. Several proposals classify order sensitive and order insensitive messages~\cite{data-dependent_ordering,sui_mysticeti,rivaseahorse,vedula2023masquerade}, but this classification has not been applied to fair ordering consensus.

Sui~\cite{sui_mysticeti} gives messages whose state changes affect only the sender a fast commit path that bypasses consensus. Applying this idea to fair ordering is promising future work. The challenge is that letting senders declare sensitivity opens a new manipulation vector. A sender could label an order sensitive transaction as insensitive to gain a latency advantage. A useful direction is to expose order sensitivity in a way that is publicly verifiable and cannot be manipulated, for example by deriving it from the read and write sets of the smart contract. Fair ordering would then apply only where it changes the outcome of execution.

\subsection{Message Ordering as a Service}
Most proposals integrate ordering into the consensus protocol to keep metadata exchange and verification overhead low. There is a growing trend, however, to separate sequencing from consensus and possibly outsource it~\cite{proposer_builder_separation,sui_consensus,sok_decentralized_sequencers,Flashbots,bearer2024espresso}, so that specialized services sequence messages before validators commit them. Distributed fair ordering could likewise become a service shared by several ledgers. This echoes Ethereum's proposed separation of block proposal from transaction ordering~\cite{proposer_builder_separation} and the modular blockchain paradigm~\cite{modular_blockchain}. To our knowledge, no work explores such an outsourced fair ordering system. The open problem is whether it can provide publicly verifiable sequencing without becoming a centralized MEV relay or a censorship chokepoint.

\section{Detailed Comparison, Overhead, and Guidance Tables}\label{sec:detailed_comparison}

This appendix collects the supporting tables referenced in the main text, namely the analytical overhead estimates (Table~\ref{tab:overhead}), the practitioner decision matrix (Table~\ref{tab:guidance}), and the per protocol FIFO and random ordering comparisons (Tables~\ref{tab:FIFO_comparison} and~\ref{tab:random_comparison}).

\begin{table}[t]
\caption{Analytical Communication and Ordering Overhead Estimates}
\label{tab:overhead}
\centering
\begin{threeparttable}
\scriptsize
\setlength{\tabcolsep}{2.5pt}
\renewcommand{\arraystretch}{1.0}
\resizebox{\textwidth}{!}{%
\begin{tabular}{@{}l|rr|rr|rr|rr@{}}
\toprule
& \multicolumn{4}{c|}{$n=10,\; m=100$} & \multicolumn{4}{c}{$n=100,\; m=1000$} \\
\cmidrule(lr){2-5}\cmidrule(l){6-9}
Protocol & Comm. & Ratio & Ord. ops & Ratio & Comm. & Ratio & Ord. ops & Ratio \\
\midrule
HotStuff (baseline) & 234.0 KB & 1.0$\times$ & 100 & 1.0$\times$ & 24.3 MB & 1.0$\times$ & 1.0K & 1.0$\times$ \\
Aequitas & 431.7 KB & 1.8$\times$ & 100.0K & 1000.0$\times$ & 440.5 MB & 18.2$\times$ & 100.0M & 100000.0$\times$ \\
Themis & 350.7 KB & 1.5$\times$ & 100.0K & 1000.0$\times$ & 308.3 MB & 12.7$\times$ & 100.0M & 100000.0$\times$ \\
Pomp\=e & 409.2 KB & 1.7$\times$ & 1.6K & 16.0$\times$ & 42.3 MB & 1.7$\times$ & 109.0K & 109.0$\times$ \\
Wendy & 409.2 KB & 1.7$\times$ & 1.6K & 16.0$\times$ & 42.3 MB & 1.7$\times$ & 109.0K & 109.0$\times$ \\
Hashgraph & 859.4 KB & 3.7$\times$ & 1.6K & 16.0$\times$ & 83.9 MB & 3.5$\times$ & 109.0K & 109.0$\times$ \\
Helix & 573.3 KB & 2.5$\times$ & 600 & 6.0$\times$ & 333.3 MB & 13.7$\times$ & 9.0K & 9.0$\times$ \\
BlindPerm & 439.1 KB & 1.9$\times$ & 100 & 1.0$\times$ & 45.5 MB & 1.9$\times$ & 1.0K & 1.0$\times$ \\
\bottomrule
\end{tabular}%
}
\begin{tablenotes}[para,flushleft]
\scriptsize
\item \textbf{Note:} Estimates are analytical, computed from protocol specifications assuming BLS signatures (96\,B), SHA-256 hashes (32\,B), and 256\,B average transaction size. Units are binary (1\,KB $= 1024$\,B). The HotStuff baseline includes full transaction broadcast in the proposal phase. ``Ratio'' is relative to vanilla HotStuff without fair ordering. Hashgraph's gossip overhead is modeled as amortized steady state cost with a $2\times$ retransmission factor. BlindPerm includes amortized DKG cost (one DKG per 10 rounds). Appendix~\ref{sec:overhead_model} gives the per protocol formulas.
\end{tablenotes}
\end{threeparttable}
\end{table}

\begin{table}[t]
\caption{Practitioner Decision Matrix for Fair Ordering Selection}
\label{tab:guidance}
\centering
\begin{threeparttable}
\scriptsize
\setlength{\tabcolsep}{2pt}
\renewcommand{\arraystretch}{1.0}
\begin{tabular}{@{}p{1.8cm}p{2.6cm}p{3.2cm}@{}}
\toprule
Scenario & Recommended Approach & Evidence Based Rationale \\
\midrule
Small permissioned committee ($n \leq 20$) requiring arrival order preservation &
FIFO via voting (Themis) &
At $n{=}10$, communication overhead is $1.5{\times}$ baseline (Table~\ref{tab:overhead}). Tolerates bounded observation falsification (Table~\ref{tab:adv_capability}). $O(m^2 \cdot n)$ ordering cost is manageable at small $n$. \\
\addlinespace
Any committee size requiring FIFO with low communication overhead &
FIFO via timestamping (Pomp\=e/Wendy) &
Communication overhead remains ${\sim}1.7{\times}$ baseline even at $n{=}100$ (Table~\ref{tab:overhead}). Requires synchronized clocks (Table~\ref{tab:FIFO_comparison}). $f$-robust guarantee holds only when honest timestamp distributions are separated (Def.~\ref{def: robust median order}). \\
\addlinespace
Large committee ($n > 50$) where arrival order preservation is not required &
Random ordering (permutation based) &
$O(m)$ ordering computation (Table~\ref{tab:random_comparison}). Immune to observation falsification and scheduling (Table~\ref{tab:adv_capability}). No clock assumptions. Guarantee is conditional on inclusion. \\
\addlinespace
Trust minimized, permissionless setting requiring public verifiability &
Blind ordering + random permutation &
Provides cryptographic content privacy and publicly auditable randomness. FIFO approaches lack public verifiability of arrival metadata. \\
\addlinespace
PBS or centralized builder architecture &
Blind ordering or TEE constrained builders &
Standard fair ordering is structurally incompatible with unregulated builders (Section~\ref{sec:pbs}). Neither path has been deployed at production scale for fair ordering. \\
\bottomrule
\end{tabular}
\begin{tablenotes}[para,flushleft]
\scriptsize
\item \textbf{Note:} All recommendations assume $n \geq 3f+1$ except FIFO via voting, which requires $n \geq 4f+1$ at $\gamma=1$. None of these approaches provides inclusion fairness; complementary mechanisms (e.g., inclusion lists) are required for end-to-end MEV protection.
\end{tablenotes}
\end{threeparttable}
\end{table}

\subsection{FIFO Ordering Protocols}

Themis~\cite{Themis_ordering} includes a comparison of FIFO protocols. Table~\ref{tab:FIFO_comparison} extends it. The table includes only proposals that give a clear fair ordering algorithm together with a fairness guarantee (see the inclusion criteria in Section~\ref{sec:scope}).

\begin{table}[t]
\caption{Comparison of FIFO Ordering Consensus Protocols Across Design, Assumptions, and Complexity}
\label{tab:FIFO_comparison}
\centering
\begin{threeparttable}
\scriptsize
\setlength{\tabcolsep}{2pt}
\renewcommand{\arraystretch}{1.0}

\resizebox{\textwidth}{!}{%
\begin{tabular}{@{}l|
  >{\RaggedRight\arraybackslash\hspace{0pt}}p{1.5cm}
  >{\RaggedRight\arraybackslash\hspace{0pt}}p{1.6cm}|
  >{\RaggedRight\arraybackslash\hspace{0pt}}p{2cm}
  >{\RaggedRight\arraybackslash\hspace{0pt}}p{1.4cm}|
  >{\RaggedRight\arraybackslash\hspace{0pt}}p{1.1cm}
  >{\RaggedRight\arraybackslash\hspace{0pt}}p{0.9cm}|
  >{\RaggedRight\arraybackslash\hspace{0pt}}p{1.6cm}
  >{\RaggedRight\arraybackslash\hspace{0pt}}p{1.6cm}@{}}
\toprule
& \multicolumn{2}{c|}{Design}
& \multicolumn{2}{c|}{Security}
& \multicolumn{2}{c|}{Assumptions}
& \multicolumn{2}{c}{Complexity} \\
\cmidrule(l{0.0em}r{0.0em}){2-3}
\cmidrule(l{0.0em}r{0.0em}){4-5}
\cmidrule(l{0.0em}r{0.0em}){6-7}
\cmidrule(l{0.0em}r{0.0em}){8-9}
\text{Proposal} & \text{Leader} & \text{Rule}
& \text{Guarantee} & \text{BFT ($n \geq$)}
& \text{Network} & \text{Clocks}
& \text{Communication} & \text{Ordering} \\
\midrule

Aequitas~\cite{voting_ordering} &
Leaderless &
Voting &
$\gamma$-batch order &
$\tfrac{2f(\gamma+1)}{2\gamma-1}$\tnote{a} &
Async. &
No &
$O(n^3)$ &
$O(m^2 \cdot n)$ \\

Quick~\cite{differential_order_fairness} &
Leaderless &
Voting &
$\gamma$-batch order &
0\tnote{b} &
Async. &
No &
$O(n^2)$ &
$O(m^2 \cdot n)$ \\

Themis~\cite{Themis_ordering} &
Leader based &
Voting &
$\gamma$-batch order &
$\tfrac{2f(\gamma+1)}{2\gamma-1}$ &
Partially sync. &
No &
$O(n^2)$ &
$O(m^2 \cdot n)$ \\

Pomp\=e~\cite{timestamping1} &
Leader based &
Timestamping &
$f$-Robust Median &
$3f+1$ &
Partially sync. &
Yes &
$O(n^2)$ &
$O(n \log n)$ \\

Wendy~\cite{timestamping_wendy} &
Leader based\tnote{c} &
Timestamping &
$f$-Robust Median &
$3f+1$ &
Async.\tnote{d} &
Yes &
$O(n^2)$ &
$O(n \log n)$ \\

Hashgraph~\cite{hashgraph} &
Leaderless &
Timestamping &
$f$-Robust Median &
$3f+1$ &
Async. &
Yes &
$O(n)$ &
$O(n \log n)$ \\

Chronos~\cite{zhang2024chronos} &
Leaderless &
Timestamping &
$f$-Robust Median &
$3f+1$ &
Async. &
Yes &
$O(n^2)$ &
$O(n \log n)$ \\

\bottomrule
\end{tabular}%
}

\begin{tablenotes}[para,flushleft]
\scriptsize
\item \textbf{Note:} $n$ is the number of nodes; $f$ is the number of Byzantine faults; $m$ is the number of messages in a round of consensus. \\
\item[a] In the case of a long Condorcet cycle, Aequitas does not provide a liveness guarantee even when all nodes are honest. \\
\item[b] Quick tolerates Byzantine faults for safety (no two honest nodes commit conflicting orders) but loses liveness with any Byzantine node: a single Byzantine participant can prevent progress. \\
\item[c] Wendy is only a pre-protocol for a consensus protocol, but this pre-protocol itself requires a leader. \\
\item[d] Also depends on the specific consensus protocol Wendy integrates with.
\end{tablenotes}
\end{threeparttable}
\end{table}

Voting based proposals have higher computational complexity because they build a dependency graph from the local orderings of all nodes. Timestamping approaches have lower ordering complexity, since their bottleneck is sorting timestamps.

Voting based proposals also tolerate fewer Byzantine faults than timestamping proposals, even at the most relaxed fairness parameter ($\gamma = 1$). Both Aequitas~\cite{voting_ordering} and Themis~\cite{Themis_ordering} tolerate up to one fourth Byzantine nodes. The reason is that the fairness condition requires $\gamma n$ nodes to agree on precedence. Since up to $f$ of them may lie, the protocol needs $\gamma n - f > \frac{n}{2}$, which limits fault tolerance.

\subsection{Random Ordering Protocols}
\begin{table}[t]
\caption{Comparison of Random Ordering Consensus Protocols Across Design, Assumptions, and Complexity}
\label{tab:random_comparison}
\centering
\begin{threeparttable}
\scriptsize 
\setlength{\tabcolsep}{2pt}
\renewcommand{\arraystretch}{1.0} 

\resizebox{\textwidth}{!}{%
\begin{tabular}{@{}l|
  >{\raggedright\arraybackslash}p{1.6cm}
  >{\raggedright\arraybackslash}p{1.6cm}|
  >{\raggedright\arraybackslash}p{2.2cm}
  >{\raggedright\arraybackslash}p{1.5cm}|
  >{\raggedright\arraybackslash}p{1.6cm}
  >{\raggedright\arraybackslash}p{0.9cm}|
  >{\raggedright\arraybackslash}p{1.8cm}
  >{\raggedright\arraybackslash}p{1.8cm}@{}}
\toprule
& \multicolumn{2}{c|}{Design}
& \multicolumn{2}{c|}{Security}
& \multicolumn{2}{c|}{Assumptions}
& \multicolumn{2}{c}{Complexity} \\
\cmidrule(l{0.0em}r{0.0em}){2-3}\cmidrule(l{0.0em}r{0.0em}){4-5}\cmidrule(l{0.0em}r{0.0em}){6-7}\cmidrule(l{0.0em}){8-9}
\text{Proposal} & \text{Leader} & \text{Rule}
& \text{Guarantee} & \text{BFT}
& \text{Network} & \text{Clocks}
& \text{Communication} & \text{Ordering} \\
\midrule

Helix~\cite{blind_fairness3_Helix} &
Leader based &
Random selection &
Probabilistic random selection &
$n \geq 3f+1$ &
Sync. &
No &
$O(n^2)$ &
$O(m \log m)$ \\

$\Pi^3$~\cite{alpos2023eating}\tnote{a} &
Leader based &
Permutation &
Permuted ordering &
N/A &
N/A &
No &
N/A &
$O(m)$ \\

MEVade~\cite{piet2023mevade}\tnote{a} &
Leader based &
Permutation &
Permuted ordering &
N/A &
N/A &
No &
N/A &
$O(m)$ \\

BlindPerm~\cite{kavousi2023blindperm} &
Leader based &
Permutation &
Permuted ordering &
$n \geq 3f+1$ &
Partially sync. &
No &
$O(n^2)$ &
$O(m)$ \\

\bottomrule
\end{tabular}%
}

\begin{tablenotes}[para,flushleft]
\scriptsize
\item \textbf{Note:} $n$ is the number of nodes; $f$ is the number of Byzantine faults; $m$ is the number of messages in a round of consensus. \\
\item[a] $\Pi^3$ and MEVade are modules that can be adapted to different leader based protocols.
\end{tablenotes}
\end{threeparttable}
\end{table}

Table~\ref{tab:random_comparison} lists four random ordering proposals. The permutation approach has the lowest ordering complexity in this paper, since it avoids both dependency graphs and timestamp sorting. It needs no synchronized clocks and integrates with various consensus protocols~\cite{alpos2023eating,piet2023mevade}. These properties make it a lightweight defense against manipulation of positions after inclusion, but not against manipulation of inclusion itself.

\section{Cost Model for Table~\ref{tab:overhead}}\label{sec:overhead_model}

We use $n$ nodes and $m$ transactions per round. Sizes are $|tx|=256$\,B for a transaction body, $|id|=32$\,B for a transaction identifier, $|h|=32$\,B for a hash, $|\sigma|=96$\,B for a BLS signature, $|\tau|=8$\,B for a timestamp, $|e|=16$\,B for a message envelope, and $|ct|=|tx|+48$\,B for an encrypted transaction. Let $c=|h|+|\sigma|+|e|$ be the size of one vote, certificate, or agreement message. Communication counts every byte sent in one round, with a message to $k$ recipients counted $k$ times. Ordering operations count the dominant comparisons or hashes per round.

\paragraph{Ordering operations.} Baseline batch assembly costs $m$. Voting based FIFO builds a pairwise preference graph in $n m^2$. Timestamping computes a median per transaction and sorts, in $nm + m\lfloor \log_2 m\rfloor$. Helix sorts hashes in $m\lfloor \log_2 m\rfloor$. BlindPerm permutes in $m$.

\paragraph{Communication.} HotStuff broadcasts the proposal and runs three phases of votes and certificates, $(n-1)(m|tx|+|\sigma|+|e|) + 3\,(2n-1)\,c$. Aequitas exchanges signed local orderings all-to-all and runs cubic agreement, $n(n-1)(m|id|+|\sigma|+|e|) + n^3 c$. Themis collects signed orderings at the leader, broadcasts them with the proposal so that replicas can verify, and runs the three HotStuff phases, $n(m|id|+|\sigma|+|e|) + (n-1)\big((n+1)m|id|+|\sigma|+|e|\big) + 3\,(2n-1)\,c$. Pomp\=e and Wendy disseminate transactions as in HotStuff, collect signed timestamps at the leader, broadcast the sorted order with timestamp proofs, and run quadratic agreement, $(n-1)(m|tx|+|\sigma|+|e|) + n\big(m(|id|+|\tau|+|\sigma|)+|e|\big) + (n-1)\big(m(|id|+|\tau|)+|\sigma|+|e|\big) + n^2 c$. Hashgraph wraps each transaction in a gossip event that reaches all $n$ nodes with a $2\times$ retransmission factor, $2nm\,(|tx|+2|h|+|\tau|+|\sigma|+|e|)$. Helix broadcasts encrypted transactions, exchanges verification hashes all-to-all, and runs quadratic agreement, $(n-1)(m|ct|+|\sigma|+|e|) + n(n-1)(m|h|+|\sigma|+|e|) + n^2 c$. BlindPerm broadcasts encrypted transactions, sends a decryption share per transaction, amortizes one DKG over ten rounds, and runs quadratic agreement, $(n-1)\,n\lfloor m/n\rfloor\,|ct| + nm(48+|\sigma|+|e|) + \lfloor n^2(48+|\sigma|+|e|)/10\rfloor + n^2 c$. The script that evaluates these formulas is included with the paper's source.

\end{document}